\documentclass[conference]{IEEEtran}
\IEEEoverridecommandlockouts
 \usepackage[utf8]{inputenc}
\usepackage{cite}
\usepackage{amsmath,amssymb,amsfonts}
\usepackage{algorithm}
\usepackage[noend]{algpseudocode}
\usepackage{graphicx}
\usepackage{tikz}
\usepackage{textcomp}
\usepackage{xspace}
\usepackage[normalem]{ulem}
\usepackage[table,xcdraw]{xcolor}

\usepackage[breaklinks=true]{hyperref}
\usepackage{tabularx} 
\usepackage{tikz}
\usetikzlibrary{shapes.geometric, arrows.meta, positioning, matrix}

\def\BibTeX{{\rm B\kern-.05em{\sc i\kern-.025em b}\kern-.08em
    T\kern-.1667em\lower.7ex\hbox{E}\kern-.125emX}}
\begin{document}

\title{AegisSat:\\ Securing AI-Enabled SoC FPGA Satellite Platforms}

\author{
  \IEEEauthorblockN{Huimin Li$^{\dagger}$,
  Vusal Novruzov$^{*}$,
  Nikhilesh Singh$^{\dagger}$,
  Lichao Wu$^{\dagger}$,
  Mohamadreza Rostami$^{\dagger}$,
  Ahmad-Reza Sadeghi$^{\dagger}$}
  
  \IEEEauthorblockA{Technische Universit\"at Darmstadt, Germany\\
  Email: $^{\dagger}$\{huimin.li, nikhilesh.singh, lichao.wu, mohamadreza.rostami, ahmad.sadeghi\}@trust.tu-darmstadt.de\\
  Email: $^{*}$\{vusal.novruzov\}@stud.tu-darmstadt.de}
}

\maketitle

\newcommand{\ourname}[0]{AegisSat}
\newcommand{\ournamewspace}[0]{\ourname\xspace}

\newcommand{\boom}[0]{BOOM\xspace}

\definecolor{spec_color}{HTML}{ff8d98}

\newcommand{\nadd}[1]{{\color{brown}{#1}}}
\newcommand{\notsure}[1]{{\color{red}{#1}}}

\newcommand{\ltodo}[1]{{\color{orange}{#1}}}

\begin{abstract}
The increasing adoption of System-on-Chip Field-Programmable Gate Arrays (SoC FPGAs) in AI-enabled satellite systems, valued for their reconfigurability and in-orbit update capabilities, introduces significant security challenges. Compromised updates can lead to performance degradation, service disruptions, or adversarial manipulation of mission outcomes. To address these risks, this paper proposes a comprehensive security framework, \ournamewspace. It ensures the integrity and resilience of satellite platforms by (i) integrating cryptographically-based secure boot mechanisms to establish a trusted computing base; (ii) enforcing strict runtime resource isolation; (iii) employing authenticated procedures for in-orbit reconfiguration and AI model updates to prevent unauthorized modifications; and (iv) providing robust rollback capabilities to recover from boot and update failures and maintain system stability. To further support our claims, we conducted experiments demonstrating the integration of these mechanisms on contemporary SoC FPGA devices. This defense-in-depth framework is crucial for space applications, where physical access is impossible and systems must operate reliably over extended periods, thereby enhancing the trustworthiness of SoC FPGA-based satellite systems and enabling secure and resilient AI operations in orbit. 
\end{abstract}

\begin{IEEEkeywords}
Satellite Security, SoC FPGA, AI, Reconfiguration, Multi-tenant.
\end{IEEEkeywords}

\section{Introduction}
\label{sec:intro}

Satellites, which used to be static platforms with fixed configurations and limited adaptability, are evolving into intelligent, reconfigurable systems capable of performing sophisticated data processing tasks in orbit~\cite{rapuano2021fpga, george2018onboard, mohamed2014development, fuchs2017bringing}. This transformation is driven by the increasing demand for enhanced autonomy, reduced dependence on ground-based control, rapid self-decision-making in mission-critical scenarios, and optimized use of constrained communication bandwidth~\cite{mohamed2014development, zhao2016reconfiguration, rapuano2021fpga}.
On the other hand, the rapid advancement of Artificial Intelligence (AI) and edge computing technologies is revolutionizing the architecture and capabilities of modern space systems~\cite{mohamed2014development, petry2024zero, petry2023accelerated, rapuano2021fpga}. AI enables critical onboard functions, including real-time data analytics, adaptive payload management, and autonomous decision-making under conditions of limited or delayed communication~\cite{petry2024zero}. These capabilities enhance the resilience, efficiency, and operational independence of satellites across diverse mission profiles~\cite{pelton2020handbook, vasudevan2020design, rapuano2021fpga}.

FPGAs have emerged as a preferable platform to meet the stringent performance requirements, adaptability, and energy efficiency in these missions~\cite{kuwahara2010fpga, leonard2025fpga, raoofyinvited}. Besides, their reconfigurability, deterministic latency, and parallel processing capabilities make them ideal for optimizing and accelerating AI workloads, such as neural network inference and real-time machine learning (ML) enhanced signal processing. Additionally, Commercial Off-The-Shelf (COTS) FPGAs are increasingly adopted in small satellites, thanks to their compact size, rapid prototyping, in-orbit updates, and flexible functionality, supporting various academic, military, and industrial applications~\cite{leonard2025fpga, langer2023robust, viel2017module, legat2012seu, eine2024reconfiguration}. 
System-on-Chip (SoC) FPGAs further enhance these advantages by integrating general-purpose processors, referred to as the \textit{Processing System} (PS), with programmable logic, known as the \textit{Programmable Logic} (PL), fabricated together within a single silicon device~\cite{kiruki2021study, rapuano2021fpga}. This hybrid architecture enables tight coupling between software control logic and hardware-accelerated AI engines, supporting complete AI processing pipelines with minimal latency and interconnect overhead. 

Unfortunately, similar to other hardware devices, satellites have been targeted by real-world cyberattacks on multiple occasions. A prominent example is the ViaSat Cyberattack during the Russo-Ukrainian War, where attackers exploited a security vulnerability from the ground segment to achieve privilege escalation~\cite{boschetti2022space, yadav2024orbital}. Indeed, the flexibility of SoC FPGAs, while enabling powerful in-orbit reconfiguration and AI acceleration, also introduces significant security challenges. A compromised configuration change or an AI model update could lead to performance degradation, service disruption, or adversarial manipulation of mission objectives. The inability to physically access satellites after launch further amplifies these risks~\cite{yadav2024orbital}. 
Despite these concerns, prior research has addressed only fragments about the security of AI-enabled SoC FPGA satellite platforms. For example, Li et al. proposed a partial reconfiguration method for satellite cryptographic devices~\cite{li2023fpga}. Cotret et al. presented lightweight reconfiguration security services for SoC FPGAs~\cite{cotret2012lightweight}, and Vallez et al. investigated onboard AI model updates while controlling their size and integrity~\cite{vallez2023efficient}. Yet, no prior work has systematically integrated these aspects into a unified framework.

Satellites of the \textit{multi-tenant model} paradigm enable multiple stakeholders, from commercial entities to government agencies, to share satellite resources such as payloads, sensors, and communication channels. By supporting diverse workloads concurrently, multi-tenant satellites improve utilization, reduce costs, and enable flexible Satellite-as-a-Service (SaaS) deployments~\cite{swartwout2016small, handley2019mega}. However, while cloud-FPGA literature has discussed partitioning programmable logic into virtual FPGAs (vFPGAs) for isolation, this concept has not been adapted to multi-tenant satellite architectures, where preventing leakage or Trojan injection between tenants is critical. Recent works, such as~\cite{exodus2020satellite, yadav2024orbital}, have highlighted the potential of multi-tenant satellite platforms but stopped short of describing concrete implementations or security strategies.

\begin{figure}
    \centering
    \includegraphics[width=\linewidth]{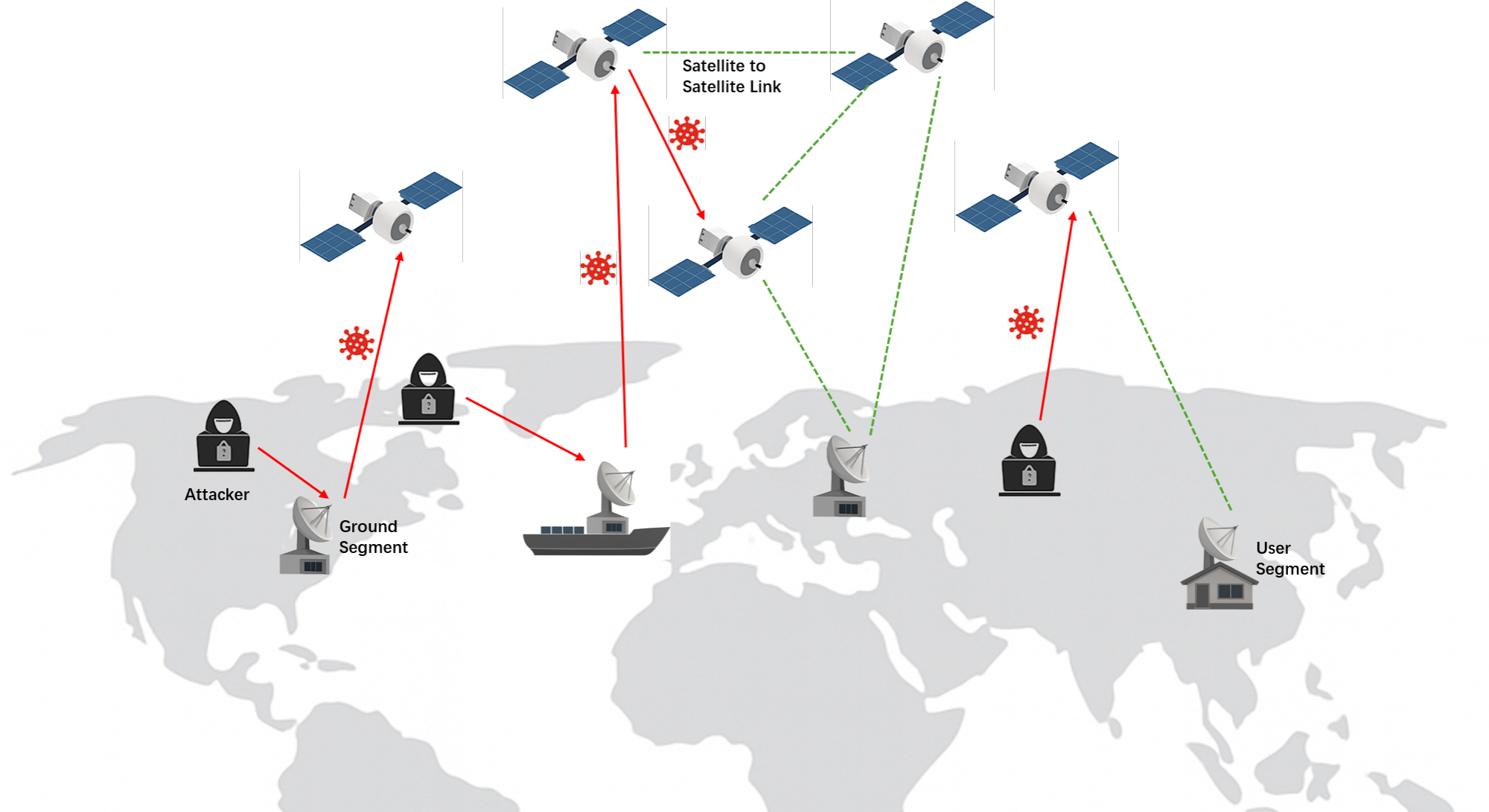}
    \caption{Overview of Satellite System and Attack Vectors. The satellite system consists of space segment, the ground segment, and the user segment. Adversaries can target the ground infrastructure or attack satellites directly. Moreover, if a single satellite is compromised, the threat can propagate to other satellites through inter-satellite communication links, highlighting the systemic risks of federated constellations.}
    \label{fig:system}
\end{figure}

Finally, as satellite-to-satellite communication becomes more prevalent, federated constellations will enable collaborative machine learning workflows, such as distributed training and in-orbit model sharing, as illustrated in Fig.~\ref{fig:system}. This evolution significantly broadens the attack surface, introducing threats that are unique to machine learning pipelines, including model injection, adversarial perturbations, and model exfiltration. These risks are further amplified when unverified updates propagate across interconnected platforms, potentially compromising the integrity and confidentiality of shared AI models~\cite{ferrag2023poisoning, barros2022novel}.

To address these challenges, this paper proposes a comprehensive security framework, named \ournamewspace\footnote{Aegis
was the shield of Zeus, meaning protection. Sat is the abbreviation for Satellite.} for AI-enabled SoC FPGA satellite platforms. Our approach establishes defense-in-depth by combining (i) cryptographically anchored secure boot, (ii) runtime isolation through TrustZone and hardware firewalls, (iii) authenticated reconfiguration and AI model updates, and (iv) robust rollback protection. Collectively, these mechanisms safeguard reconfigurable satellite platforms throughout their operational lifecycle in the inaccessible environment of space. 
Our contributions are as follows:
\begin{itemize}
\item We develop a systematic threat model that characterizes the attack surfaces resulting from the convergence of reconfigurable logic, AI acceleration pipelines, and multi-tenant mission architectures.
\item We propose a layered security framework, \ournamewspace that unifies secure initialization, runtime protection, and lifecycle update mechanisms into an end-to-end defense architecture.
\item We conduct experiments demonstrating the integration of these mechanisms on contemporary SoC FPGA devices.
\item We outline a forward-looking research agenda highlighting critical open challenges to counter emerging threats.
\end{itemize}

\section{Background}
\label{sec:background}

\begin{table*}[htbp]
\centering
\caption{Survey of SoC FPGA-based AI Implementations for aerospace Applications.}
\label{tab:soc_fpga_ai_space}
\resizebox{\textwidth}{!}{%
\begin{tabular}{|l|c|c|c|c|l|}
\hline
\textbf{Authors} & \textbf{Year} & \textbf{AI Algorithms} & \textbf{SoC FPGA} & \textbf{Vendor} & \textbf{Applications} \\ \hline
Pitsis et al.~\cite{pitsis2019efficient} & 2019 & CNN & Zynq UltraScale+ MPSoC & AMD Xilinx & Space data classification \\ \hline
Ma et al.~\cite{ma2019lightweight, antunes2025fpga} & 2019 & Autoencoder NN & Zynq UltraScale+ MPSoC & AMD Xilinx & Feature extraction and anomaly detection \\ \hline
Sabogal et al.~\cite{sabogal2019recon} & 2019 & CNN & Zynq UltraScale+ MPSoC & AMD Xilinx & Semantic segmentation of space imagery \\ \hline
Liu et al.~\cite{liu2019towards} & 2019 & CNN & Arria 10 SX SoC & Intel Altera & Remote sensing image segmentation \\ \hline
Li et al.~\cite{li2019efficient} & 2019 & SSD NN & Zynq 7000 & AMD Xilinx & Remote sensing imagery analysis \\ \hline
Reiter et al.~\cite{reiter2020fpga} & 2020 & BNN & Zynq 7000 & AMD Xilinx & Real-time cloud detection \\ \hline
Lemaire et al.~\cite{lemaire2020fpga} & 2020 & BNN + CNN & Cyclone V & Intel Altera & Cloud classification and detection \\ \hline
Lent et al.~\cite{lent2020evaluating} & 2020 & SNN & Zynq 7020 & AMD Xilinx & Routing in space networks \\ \hline
Cosmas et al.~\cite{cosmas2020utilization} & 2020 & CNN & Zynq UltraScale+ MPSoC & AMD Xilinx & Visual landmark recognition for navigation \\ \hline
Zhang et al.~\cite{zhang2021fpga} & 2021 & YOLOv2 (CNN) & Zynq 7000 & AMD Xilinx & Optical object detection \\ \hline
Rapuano et al.~\cite{rapuano2021fpga} & 2021 & CNN & Zynq UltraScale+ MPSoC & AMD Xilinx & Cloud detection \\ \hline
Sabogal et al.~\cite{sabogal2021reconfigurable} & 2021 & CNN & Zynq 7020, Zynq UltraScale+ MPSoC & AMD Xilinx & Semantic segmentation \\ \hline
Pacini et al.~\cite{pacini2021multi} & 2021 & CNN & Zynq UltraScale+ MPSoC & AMD Xilinx & Real-time image classification \\ \hline
Pitonak et al.~\cite{pitonak2022cloudsatnet} & 2022 & CNN & Zynq 7020 & AMD Xilinx & Cloud detection \\ \hline
Zhang et al.~\cite{zhang2022accurate} & 2022 & GNN & Zynq UltraScale+ MPSoC & AMD Xilinx & SAR image classification \\ \hline
Abderrahmane et al.~\cite{abderrahmane2022spleat} & 2022 & SNN & Cyclone V & Intel Altera & Cloud detection \\ \hline
Papatheofanous et al.~\cite{papatheofanous2022soc} & 2022 & CNN & Zynq UltraScale+ MPSoC & AMD Xilinx & Satellite image segmentation \\ \hline
Perryman et al.~\cite{perryman2023evaluation} & 2023 & MobileNetV1, ResNet-50, GoogLeNet & XCVC1902 (VCK190) & AMD Xilinx & Edge computing in space \\ \hline
Ekblad et al.~\cite{ekblad2023resource} & 2023 & YOLOv4-based NN & Zynq UltraScale+ MPSoC & AMD Xilinx & Autonomous navigation \\ \hline
Gao et al.~\cite{gao2023systematic} & 2023 & CNN & Zynq 7000 & AMD Xilinx & CNN reliability evaluation \\ \hline
Carmeli et al.~\cite{carmeli2023ai} & 2023 & SOM NN & Cyclone V & Intel Altera & Star pattern recognition \\ \hline
Coca et al.~\cite{coca2023fpga} & 2023 & ResNet & Zynq UltraScale+ MPSoC & AMD Xilinx & Burned area anomaly detection \\ \hline
Zhao et al.~\cite{zhao2023hardware} & 2023 & YOLOv4-MobileNetv3 & Zynq UltraScale+ MPSoC & AMD Xilinx & Object detection in satellite images \\ \hline
Mazouz et al.~\cite{mazouz2024online} & 2024 & YOLOv3 & Zynq 7100 & AMD Xilinx & Streaming object detection \\ \hline
Kim et al.~\cite{kim2024fpga} & 2024 & Reinforcement Learning & Zynq 7000 & AMD Xilinx & Routing in LEO networks \\ \hline
Castelino et al.~\cite{castelino2024energy} & 2024 & Conv. Autoencoder & Zynq UltraScale+ MPSoC & AMD Xilinx & HSI artifact detection \\ \hline
Zhang et al.~\cite{zhang2024energy} & 2024 & Dehazing NN & Zynq 7000 & AMD Xilinx & Image dehazing \\ \hline
Cratere et al.~\cite{cratere2025efficient} & 2024 & CNN & Zynq UltraScale+ MPSoC & AMD Xilinx & Cloud detection \\ \hline
Kim et al.~\cite{kim2024orbit} & 2024 & SqueezeNet & Zynq 7000 & AMD Xilinx & Cloud detection \\ \hline
Li et al.~\cite{li2024fpga} & 2024 & CNN & Zynq UltraScale+ MPSoC & AMD Xilinx & Depth estimation in spacecraft \\ \hline
Upadhyay et al.~\cite{upadhyay2024design} & 2024 & ResNetc & Zynq UltraScale+ MPSoC & AMD Xilinx & Cloud detection \\ \hline
Posso et al.~\cite{posso2024real} & 2024 & Mobile-URSONet & Zynq UltraScale+ MPSoC & AMD Xilinx & Pose estimation \\ \hline
Ciancarelli et al.~\cite{ciancarelli2024special} & 2024 & Autoencoders, CNNs & Xilinx ACAP & AMD Xilinx & Anomaly detection, SAR, RF \\ \hline
Leon et al.~\cite{leon2024mpai} & 2024 & UrsoNet, MobileNetV2, ResNet-50 & Zynq UltraScale+ MPSoC & AMD Xilinx & Pose estimation and benchmarking \\ \hline
Barnwal et al.~\cite{barnwal2024morphological} & 2024 & CNN & Zynq UltraScale+ MPSoC & AMD Xilinx & Galaxy classification \\ \hline
Bai et al.~\cite{bai2024design} & 2024 & CNN & Zynq 7020 & AMD Xilinx & Particle identification \\ \hline
Jiang et al.~\cite{jiang2024efficient} & 2024 & DNN & Zynq UltraScale+ MPSoC & AMD Xilinx & Hyperspectral anomaly detection \\ \hline
Shi et al.~\cite{shi2024aircraft} & 2024 & CNN & Zynq-7000, UltraScale+ MPSoC & AMD Xilinx & Image classification \\ \hline
Justo et al.~\cite{justo2024hyperspectral} & 2024 & CNN & Zynq 7030 & AMD Xilinx & Hyperspectral segmentation \\ \hline
Renaut et al.~\cite{renaut2025deep} & 2025 & DNN & Zynq 7000 & AMD Xilinx & Satellite pose estimation \\ \hline
Garcés-Socarrás et al.~\cite{garces2025artificial} & 2025 & CNN & VC190, Zynq UltraScale+ MPSoC & AMD Xilinx & Payload config and beamforming \\ \hline
Perryman et al.~\cite{perryman2025dependable} & 2025 & CNN & XCVC1902 (VCK190), XCVE2802 (VEK280) & AMD Xilinx & Fault-tolerant AI acceleration \\ \hline
\end{tabular}%
}
\end{table*}

\subsection{AI Acceleration and Design Methodologies on SoC FPGAs}
AI acceleration on SoC FPGAs leverages the inherent parallelism of FPGA fabric to optimize the execution of machine learning algorithms. Advanced design methodologies, such as high-level synthesis (HLS) and hardware/software (HW/SW) co-design, facilitate the efficient implementation of AI workloads on these reconfigurable platforms. Neural network inference, particularly with quantized models, can be effectively mapped onto FPGA logic using HLS tools and dedicated AI model compilers \cite{leonard2025fpga, petry2023accelerated}. 

A range of toolchains supports this development process, including Xilinx Vitis AI\footnote{\url{https://github.com/Xilinx/Vitis-AI}}, Intel’s OpenVINO\footnote{\url{https://github.com/openvinotoolkit/openvino}}, hls4ml\footnote{\url{https://github.com/fastmachinelearning/hls4ml}}, FINN\footnote{\url{https://finn.readthedocs.io/en/latest/}}, and MATLAB’s HDL Coder\footnote{\url{https://www.mathworks.com/products/hdl-coder.html}}. These tools enable systematic model partitioning, quantization, and deployment tailored to the underlying FPGA architecture. %
Moreover, partial reconfiguration (PR) facilitates dynamic updates to AI accelerators at runtime, ensuring uninterrupted system functionality. Using Isolation Design Flow (IDF), designers can maintain functional and security isolation between static and dynamic regions of the hardware, supporting safe and adaptive mission reconfiguration \cite{sabogal2021reconfigurable}. Collectively, these approaches reduce development cycles, enhance design flexibility, and improve the power efficiency of AI-enabled satellite systems.

\subsection{AI in Space Applications on SoC FPGA}
Table~\ref{tab:soc_fpga_ai_space} presents a comprehensive list of recent studies in satellite missions.
Interestingly, these studies indicate a growing adoption of SoC FPGA platforms in space missions. Convolutional neural networks (CNN) are the most commonly deployed models, primarily due to their effectiveness in vision-based classification tasks. These are followed by autoencoders, spiking neural networks (SNNs), reinforcement learning algorithms, and graph neural networks (GNNs). Among the available platforms, AMD Xilinx solutions, particularly the Zynq UltraScale+ MPSoC, are the most prevalent, owing to their high reconfigurability, mature development ecosystem, and demonstrated reliability in space applications~\cite{papatheofanous2022soc}. However, the combination of reconfigurable fabrics and their supporting software stacks, including FPGA bitstream managers, partial reconfiguration workflows, and AI runtime environments, introduces complex dependencies and attack surfaces that conventional software-only security measures are not ready to defend. While prior research has focused primarily on improving inference throughput and energy efficiency, none of the works listed in Table~\ref{tab:soc_fpga_ai_space} systematically address implementation-level platform security. This gap underscores the urgent need for cohesive, hardware-assisted security measures to ensure long-term system integrity and resilience in the unique operational context of space.

\subsection{Multi-Tenancy on SoC FPGA}
\begin{figure}
    \centering
    \includegraphics[width=\linewidth]{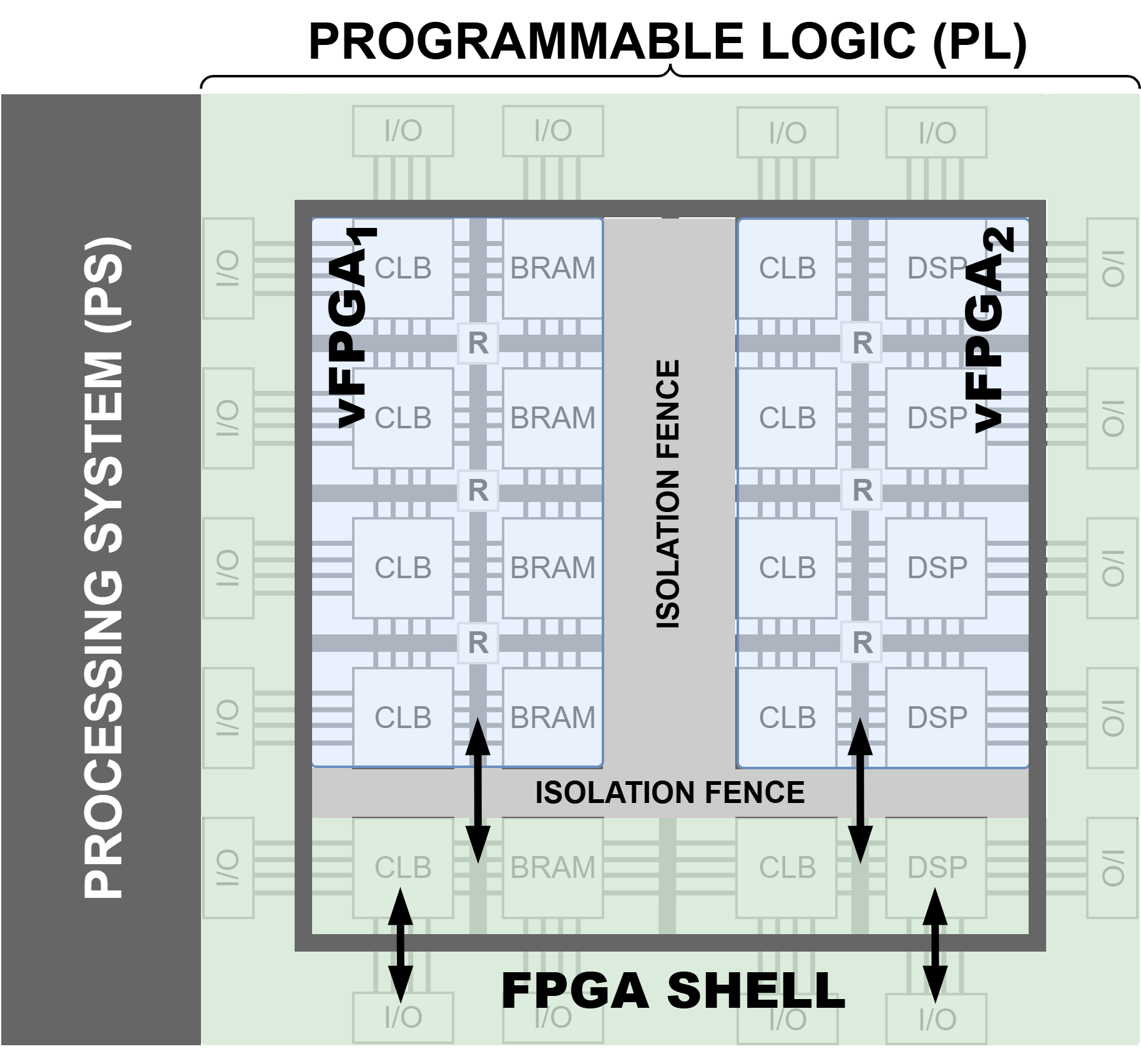}
    \caption{Multi-Tenancy on SoC FPGA. CLB: Configurable Logic Block; R: Configurable Routing Block; BRAM: Block Random Access Memory; DSP: Digital Signal Processing; I/O: Input/Output Interface. }
    \label{fig:multi-tenant}
\end{figure}

In the context of SoC FPGAs, multi-tenancy refers to the partitioning of FPGA fabric into isolated regions, each dedicated to a different tenant~\cite{dessouky2021sok}. As illustrated in Figure~\ref{fig:multi-tenant}, the PL part is partitioned into multiple reconfigurable regions (e.g., \textit{vFPGA\textsubscript{1}} and \textit{vFPGA\textsubscript{2}}), each capable of hosting isolated workloads from different tenants. These regions are managed and interfaced through a common \textit{FPGA Shell}, which provides shared infrastructure and supports secure communication with PS. This architecture enables spatial isolation and dynamic partial reconfiguration of the programmable logic, allowing independent deployment and runtime updates of distinct AI or mission workloads within a single FPGA device, an essential capability for multi-payload and adaptive satellite missions. Leveraging PR, different applications can dynamically share FPGA resources without interference~\cite{koch2012partial, karteris2024towards}. This flexibility is essential for long-duration missions, enabling the hardware to adapt to evolving requirements such as new AI models, new instruments, or protocols over time~\cite{esa2023reprogrammable}.

Multi-tenancy enhances resource efficiency, which is especially critical for constrained platforms like CubeSats. However, it also introduces significant security challenges due to the shared nature of the hardware~\cite{meda2022secure}. Potential risks include unauthorized access to a tenant’s data or configuration, data leakage between tenants, and interference through side-channel attacks, such as those exploiting timing or power consumption~\cite{dessouky2021sok}. These vulnerabilities are particularly critical in space-based systems, where adversarial environments and remote reconfiguration capabilities heighten the risk of attacks like bitstream tampering or hardware trojans~\cite{elnaggar2019multi}. 

\subsection{Security Challenges in Space-Based Systems}
Space-based platforms, particularly those leveraging reconfigurable computing architectures such as SoC FPGAs, face complex and mission-critical security challenges. The distinctive constraints of the space environment, including radiation exposure, limited physical access, and intermittent communication with ground stations, exacerbate risks from both conventional and domain-specific cyber threats. The integration of AI accelerators, support for dynamic reconfiguration, and adoption of multi-tenant processing models further broaden the attack surface, demanding comprehensive end-to-end protection mechanisms.

This paper focuses on threats to platform integrity, especially those targeting the hardware reconfiguration pipeline, AI co-processing engines, and the overall system lifecycle. Key categories of threats include:
\begin{itemize}
    \item \textbf{Fault-Induced Attacks:} Space radiation can cause Single-Event Upsets (SEUs) that alter configuration memory or computation logic. Deliberate fault injection (e.g., laser, EM pulses) can exploit similar vulnerabilities for code corruption or security bypass~\cite{campola2019radiation, ren2024overview}.  
    \item \textbf{Supply Chain Attacks:} Adversaries may insert malicious modifications during Integrated Circuit  (IC) fabrication, IP core integration, or third-party toolchain use. Hardware Trojans, designed to remain dormant during functional verification, may activate under specific triggers to exfiltrate data or disrupt operations~\cite{bhunia2014hardware, elnaggar2019multi}.
    \item \textbf{Software Exploits:} Embedded Linux and Real-Time Operating System (RTOS) platforms on SoC FPGAs often include complex drivers, middleware, and AI toolchains. Vulnerabilities in these layers may enable unauthorized access to the programmable logic or privileged control interfaces~\cite{rahman2024comprehensive, jero2024securing}. 
    \item \textbf{Reverse Engineering:} Through remote side channels, adversaries can extract proprietary bitstreams or reconstruct deployed AI model parameters, jeopardizing intellectual property and model confidentiality~\cite{zhang2019recent, tramer2016stealing}.
\end{itemize}

Beyond general threat vectors, the secure operation of SoC FPGA-based satellites critically depends on defending against low-level, architecture-specific attacks. These vulnerabilities often arise at the intersection of software, firmware, and reconfigurable hardware, providing opportunities to compromise execution integrity, circumvent isolation mechanisms, or subvert critical AI functionalities. Addressing these foundational weaknesses is essential to maintain long-term trustworthiness in autonomous and remote satellite deployments.
\begin{itemize}
    \item \textbf{Bitstream and AI Model Integrity:} Inadequate authentication of partial reconfiguration (PR) modules or AI model updates can allow injection of unauthorized logic. Adversaries may exploit this to embed covert computation units or poison AI inference results~\cite{xing2025towards}.  
    \item \textbf{Hardware Trojanized Logic:} Third-party IP cores, if unchecked, may introduce hardware Trojans that evade static analysis and activate under rare conditions. These can undermine the system’s trust boundary and persist undetected in deployed platforms~\cite{hu2020overview, hoque2020ip, giuffrida2024foss}. 
    \item \textbf{Software Stack Exploits:} Improper memory protection or weak isolation between software layers may lead to privilege escalation, control flow hijacking, or logic reprogramming. This is particularly severe in embedded SoCs where software-hardware coupling is tight~\cite{rajendran2024exploring}.
\end{itemize}

This work comprehensively examines all the above threat landscapes, with an emphasis on the interdependencies between reconfigurable hardware, embedded software, and AI workloads in space-based systems. %

\subsection{Cross-Layer and Lifecycle Security Requirements}
Ensuring the secure operation of AI-enabled SoC FPGA-based satellites requires a cross-layer security architecture covering the entire system lifecycle, from fabrication and provisioning to in-orbit updates and decommissioning. These systems face adversarial and resource-constrained environments, demanding tightly integrated hardware-software protections beyond standard embedded security.

A foundational element is the secure boot process, which uses cryptographic signatures to ensure firmware authenticity and integrity at startup, establishing a hardware-rooted chain of trust~\cite{amacher2019performance}. Runtime protections depend on isolation technologies such as ARM TrustZone, AXI firewalls, and Memory Protection Units (MPUs), which segment system components and restrict privilege escalation~\cite{amacher2019performance}.
To support long-term missions, in-orbit platforms require secure firmware and AI model updates with cryptographic signing and rollback prevention, especially under connectivity delays~\cite{sunter2016firmware, marchand2023firmware}. Meanwhile, physical-layer threats like radiation-induced Single Event Upsets (SEUs) necessitate resilience mechanisms such as  Error-Correcting Codes (ECC), Triple Modular Redundancy (TMR), or dynamic reconfiguration to preserve system reliability~\cite{ yue2022security, shrivastwa2023enhancements, diana2024review}.

\section{Threat Model}
\label{sec:sec_model}
This paper adopts a comprehensive threat model reflecting the risks faced by AI-enabled SoC FPGA-based satellites in shared and adversarial environments. We assume the attacker has full knowledge of the satellite's hardware architecture, software stack, and operational workflows. Furthermore, the attacker is granted legitimate access via satellite-as-a-service (SaaS) interfaces, allowing them to upload custom FPGA bitstreams or AI models onto shared computational payloads. Although other satellite applications are assumed benign, they are not trusted by default. Exploitation may arise from software bugs, weak isolation between the PS and PL, or insufficient runtime authentication. The attacker’s objective may include unauthorized data access, disruption of mission-critical operations, or manipulation of AI inference outcomes.

\section{Security Framework}
\label{sec:sec_frame}
Our security framework, \ournamewspace, adopts a layered architecture that integrates secure initialization, continuous runtime protection, and authenticated lifecycle management into a cohesive system. Rather than treating these components as isolated point solutions, our approach explicitly connects each stage of the platform's operation: secure boot establishes a hardware-rooted chain of trust that provisions cryptographic keys and baseline integrity measurements (\ref{sec:Secure_Boot}); trusted execution environments maintain this trust throughout runtime by isolating sensitive assets and enforcing least-privilege execution (\ref{sec:trust_exe}); and secure update workflows extend trust into the system's evolution by validating new configurations and AI models before activation (\ref{sec:secure_update}). 
During failed boot or updates, \ournamewspace provides robust fallback mechanisms that restore a known-good state from a golden image, ensuring service continuity and preventing persistent compromise.
Together, these pillars form a defense-in-depth strategy, in which each layer reinforces and complements the others to ensure resilience against compromise, even in the absence of physical access or timely intervention.

\subsection{Secure Boot and Root of Trust for SoC FPGA Satellites}
\label{sec:Secure_Boot}
In space applications, where systems must operate reliably for extended periods without physical intervention, a robust, secure boot is essential to ensure that only authenticated and integrity-verified firmware, bitstreams, and AI models are executed onboard. This safeguards mission-critical operations against unauthorized modifications and cyber threats \cite{XilinxSecureBoot}.

\subsubsection{Boot Sequence and Chain of Trust}

Secure boot in SoC FPGAs follows a hierarchical chain-of-trust model. The process begins with an immutable Boot ROM embedded in the silicon, which authenticates the first-stage bootloader (FSBL) using cryptographic algorithms such as RSA-4096 and SHA-3/384~\cite{intel_secure_boot, xapp_secure_boot}. The FSBL initializes system components and verifies each subsequent stage, including loading the operating system and configuring the programmable logic fabric~\cite{microsemi_secure_boot}. Cryptographic signatures and optional encryption are applied to software and bitstreams to ensure integrity and prevent tampering. %

\subsubsection{Cryptographic Key Management}

Robust key management is essential for maintaining the integrity of the secure boot process. SoC FPGAs like the Zynq UltraScale+ MPSoC integrate mechanisms such as eFUSE arrays for storing public key hashes~\cite{Xilinx_Zynq7000APSoCSecurity, intel_stratix10_config}. eFUSE arrays are one-time programmable memory for permanent storage of sensitive data. Battery-backed RAM (BBRAM) is used for volatile key storage, which can be cleared in response to tampering. Physical Unclonable Functions (PUFs) enhance security by deriving keys from intrinsic hardware properties, eliminating the need for persistent storage~\cite{PUFTech}. 
Tamper detection circuits erase sensitive keys if anomalies in environmental parameters are detected, ensuring security throughout the satellite's lifecycle. %

\subsubsection{Bitstream Authentication and Encryption}
To protect FPGA configurations, bitstreams are encrypted with AES-256 and authenticated using RSA or ECDSA signatures~\cite{XilinxUltraScale, intel_s10_security}. AES-256 ensures strong symmetric encryption, while RSA and ECDSA verify digital signatures. In Xilinx Zynq UltraScale+ MPSoCs, the Configuration Security Unit (CSU) manages decryption and authentication during secure boot~\cite{xilinx_ug1161}. Intel’s Stratix 10 and Agilex devices use a Secure Device Manager (SDM) with dedicated hardware engines for similar protections~\cite{intel_agilex7_soc, intel_sdm_wp01252}. To prevent replay attacks and unauthorized downgrades, modern FPGAs implement version control and anti-rollback mechanisms that compare stored version numbers against incoming bitstreams~\cite{intel_pac_security}.

\subsubsection{Failure Handling and Recovery}
In case of boot failure, SoC FPGAs employ fallback mechanisms such as a non-overwritable golden image, retry logic for alternate configurations or safe mode, and watchdog timers for system recovery. A golden image is a pre-validated, secure configuration stored in protected memory. These methods are essential for space applications, where real-time intervention is constrained and environmental variability is high~\cite{dzemaili2021reliable, siegle2016fault, la2022study}.

\subsection{Trusted Execution and Hardware Isolation}
\label{sec:trust_exe}
While secure boot establishes trust at system startup, the dynamic nature of AI-enabled satellite platforms requires continuous runtime protection~\cite{diana2024review, gross2022breaking}. This section examines the runtime threat landscape in SoC FPGA-based satellites and presents key isolation mechanisms for least-privilege execution.

\subsubsection{Runtime Threat Landscape}
During in-orbit operation, vulnerabilities may arise beyond secure boot protections. Attackers could exploit flaws in AI inference engines \cite{diana2024review, rech2024artificial, wang2024case, li2023flairs}, OS services \cite{cratere2024board}, or communication protocols \cite{yue2023low} to execute arbitrary code or escalate privileges. Malicious bitstreams loaded during runtime reconfiguration might include unauthorized logic capable of memory access or bus manipulation \cite{oche2021applications}. The tight coupling between PS and PL could allow adversaries to traverse domains, causing data leakage, denial-of-service (DOS), or logic corruption \cite{diro2025space, utsash2024implementing, wu2023comprehensive, geist2023nasa}.

\subsubsection{Execution Isolation in the Processing System}
The ARM TrustZone architecture partitions execution into Secure and Normal Worlds \cite{pinto2019demystifying}. TrustZone is a security extension that enables secure execution of critical tasks (e.g., key management, firmware validation) in the Secure World, while general-purpose processes (e.g., AI inference, data handling) run in the Normal World \cite{zhang2024enhancing}. TrustZone ensures memory and peripheral isolation and supports context switching via Secure Monitor Calls (SMCs) \cite{islam2024confidential, jian2025smartzone}. Memory Protection Units (MPUs) and Memory Management Units (MMUs) provide process-level access control, protecting AI models, telemetry data, and kernel-space buffers \cite{huang2024sok}.

\subsubsection{Isolation of Programmable Logic}
The PL fabric interfaces with the PS often through Advanced eXtensible Interface (AXI) buses, playing a critical role in system-level communication. However, if compromised, the PL can become a conduit for unauthorized access to protected memory regions or peripheral devices. To mitigate such risks, platforms such as Xilinx offer hardware enforcements like the AXI Firewall IP~\cite{xilinxaxi} and the System Memory Management Unit (SMMU) \cite{gross2019breaking, yarza2022safety}. Other vendors, such as Intel, integrate similar security primitives in their FPGA SoCs, including configurable memory protection controllers and isolation-enabled bus interconnects, to ensure secure communication boundaries between hardware and software domains~\cite{proulx2023survey, intel_s10_security, intel_sdm_wp01252}. A robust security design also necessitates validating interrupt lines originating from the PL to the PS, especially when interfacing with high-risk modules like non-volatile storage or command subsystems~\cite{christoforakis2020protection, silitonga2020mites}.

\subsubsection{Secure AI Co-Processing}
AI workloads deployed on SoC FPGAs often span both the PS and PL domains, requiring secure data exchange. Secure co-processing frameworks adopt cryptographic techniques such as Secure Hash Algorithms (e.g., SHA-256) to validate the integrity of AI computations. These hashes help ensure that data or intermediate results have not been tampered with during execution \cite{CryptoHash, li2023flairs}. Furthermore, runtime security monitors and HW/SW co-designed attestation schemes are employed to detect anomalous behaviors such as timing violations, unexpected control flow changes, or unauthorized access attempts~\cite{solet2018hw}. These mechanisms form the backbone of a zero-trust execution environment, wherein no component, whether in the PS or PL, is inherently trusted without continuous verification.

\subsubsection{Privilege Separation and Containment}
Privilege separation limits the impact of compromises. In SoC FPGAs, the bootloader validates the launch chain, TrustZone manages cryptographic assets, and AI processes run in sandboxed environments. PL modules are restricted by memory and interrupt controls~\cite{yarza2022safety}. Localized resets or reconfigurations can preserve mission continuity if a subsystem (e.g., AI accelerator) fails, aided by watchdog timers and fault monitors \cite{fons2017modular}.

\subsection{Secure Update, Partial Reconfiguration, and AI Model Updates}
\label{sec:secure_update}
While runtime isolation ensures that malicious code cannot compromise critical functions, secure and resilient in-orbit updates are essential for the adaptability and longevity of AI-driven satellite platforms. While enabling on-the-fly improvements to mission logic and AI models~\cite{vallez2023efficient}, these capabilities also expose critical attack surfaces~\cite{singh2023application, thangavel2024artificial}. The following section describes how authenticated update workflows extend and preserve trust throughout the system's lifecycle.

\subsubsection{Secure Update Workflow and Threat Mitigation}
Update commands, including logic bitstreams and AI model parameters, are transmitted via mutually authenticated and encrypted Telemetry, Tracking, and Command (TT\&C) links~\cite{bader2024requirements}. %
TT\&C links are the communication channels between ground stations and satellites. Commands include cryptographic signatures, sequence numbers, and timestamps and etc. to ensure integrity, prevent replay, and validate freshness~\cite{vasudevan2020design, morrison2006secure}.
Upon receipt, updates are stored in protected memory and cryptographically validated before deployment~\cite{mody2019ttc}. Validation failures trigger automatic discards and logged alerts. Acknowledgment protocols and retry mechanisms address transient transmission failures~\cite{toubi2024vulnerability}. This workflow mitigates threats such as bitstream injection, rollback attacks, and logic hijacking by enforcing staged validation and strict access control.

\subsubsection{Redundancy and Recovery Mechanisms}
To prevent mission degradation from failed updates, satellites retain a golden image, an immutable configuration stored in secure non-volatile memory~\cite{pinchas2009satellite, GoldenImage}. Watchdog timers monitor update behavior, reverting to the golden image upon detecting instability~\cite{cutler2022watchdog, Dong2001DesignOA}. Some architectures use dual-image buffers or redundant logic blocks for test deployments. Post-update, built-in self-tests (BIST) and output verification against golden baselines confirm operational correctness~\cite{el2007improved}.

\subsubsection{Runtime Detection of Malicious Logic}
Runtime detection of malicious logic is critical for maintaining satellite security. Techniques include monitoring system behavior for anomalies, using intrusion detection systems, and leveraging hardware-based security features~\cite{wiatrek2024advancing, rahmatian2012adaptable}. Anomaly detection algorithms identify deviations from normal operation~\cite{lu2015analysis, abbas2017power}, while hardware security modules (HSMs) protect sensitive operations. These measures ensure timely detection and mitigation of unauthorized logic or tampering~\cite{bailin2010application}.

\subsubsection{Partial Reconfiguration of FPGA Logic}
Partial Reconfiguration (PR) allows selective updates to FPGA logic without interrupting the rest of the system~\cite{wankhade2014dynamic, burman2017development}. For AI workloads, PR enables modular accelerator upgrades without full reboots~\cite{manjith2017adaptive}. Static and dynamic regions are defined using tools like Xilinx’s Isolation Design Flow, enforcing strict boundaries. PR bitstreams are authenticated and decrypted using AES-256 and RSA/ECDSA before activation~\cite{unterstein2020secure}. PR modules are sandboxed upon load, with behavior constrained by memory access restrictions and interface isolation. Validation with test vectors and integrity checks precede operational use~\cite{wankhade2014performance}. A trusted controller in the static region orchestrates PR operations, ensuring traceability and mitigating logic-level attacks.

\subsubsection{AI Model Lifecycle and Security}
AI models must adapt to evolving tasks and environments. Software-based models are stored in encrypted memory and verified during runtime load~\cite{vallez2023efficient}. Hardware-accelerated models (e.g., quantized neural networks) receive updates via PR or secure memory transactions. To prevent model poisoning~\cite{anan2024securing, ModelPoisoning}, updates are cryptographically signed and checked against expected behavior. Secure loading is complemented by functional testing, and in federated constellations, model propagation is governed by consensus or multi-signature authorization~\cite{ferrag2023poisoning, barros2022novel}.

\subsubsection{Operational and Scheduling Considerations}
In-orbit updates must align with operational safety and mission constraints. Updates are scheduled during idle periods with stable thermal and power profiles~\cite{sakib2023orbit}. Configuration clocks must remain within tolerance, and critical maneuvers are paused during reconfiguration~\cite{ren2022orbit}. Failure handling includes retry logic, Cyclic Redundancy Check (CRC) validation, and rollback triggers~\cite{hanafi2022fail}. A hardened PR controller ensures deterministic sequencing~\cite{cao2024orbit}, while event logs support diagnostics and forensics~\cite{zhao2024intelligent}.

\section{Implementation}
\label{sec:impl}
To demonstrate the feasibility of the proposed security mechanisms, we implemented a proof-of-concept on the Xilinx ZCU102 development board. This platform integrates the Xilinx Zynq UltraScale+ MPSoC, which combines a quad-core ARM Cortex-A53 (with TrustZone), dual Cortex-R5 real-time cores, and FPGA programmable logic within a single chip. The hardware setup was configured and managed using Xilinx Vitis 2023.1. 
The device features dedicated secure boot capabilities via the Configuration Security Unit (CSU), hardware support for AES-256 encrypted bitstreams and RSA authentication, and TrustZone-based isolation for secure processing.
During system provisioning, the eFUSE arrays were programmed with an RSA public key hash to enable cryptographic verification of all boot components. Secure boot was activated to ensure only authenticated firmware and FPGA bitstreams can be loaded.

\begin{figure}
    \centering
    \includegraphics[width=\linewidth]{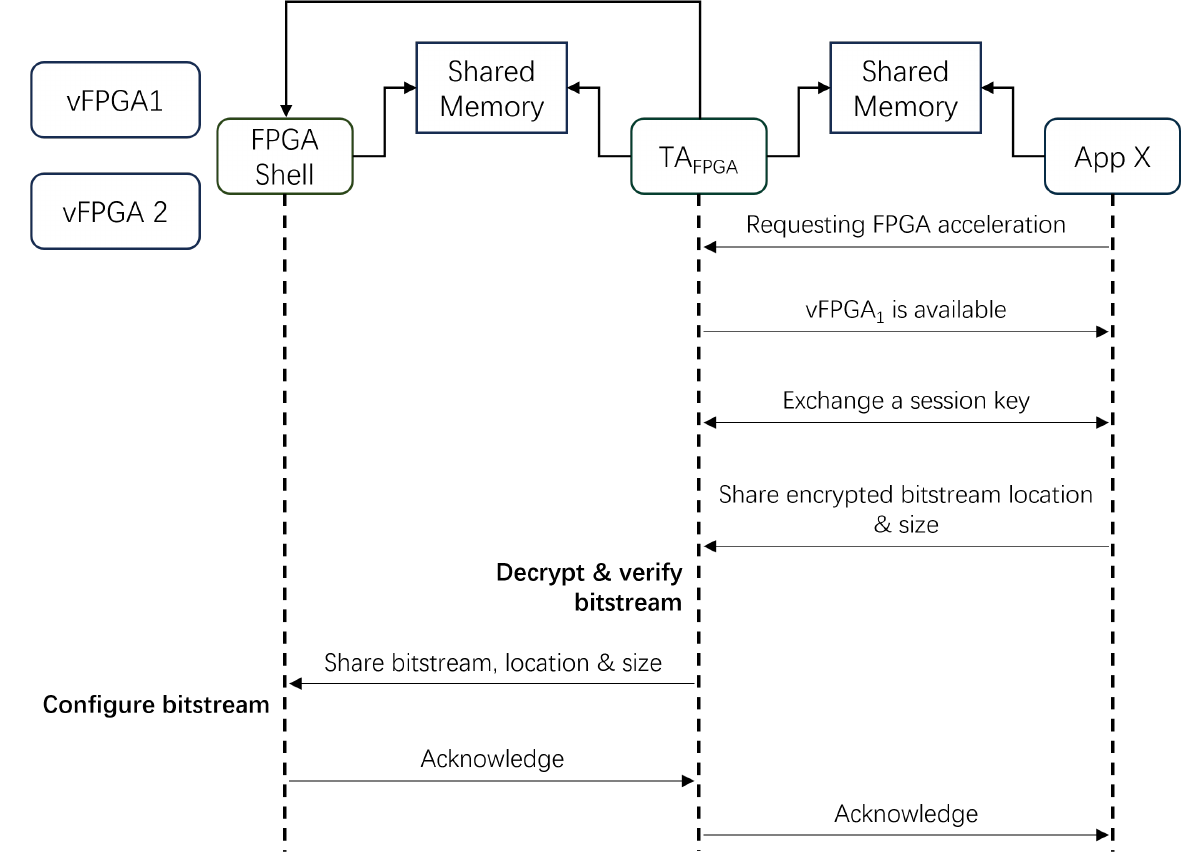}
    \caption{Secure Reconfiguration Workflow.}
    \label{fig:workflow}
\end{figure}

The PL part was partitioned into three fixed regions: one static region hosting the FPGA shell and two dynamically reconfigurable virtual FPGA (vFPGA) partitions allocated to user applications. Each vFPGA was assigned dedicated I/O interfaces and unique memory address ranges. This setup allows independent management, configuration, and isolation of workloads. 
A trusted application on the PS part, referred to as the FPGA Trust Anchor ($TA_{\text{FPGA}}$), was implemented to control bitstream decryption, validation, and reconfiguration. 
The $TA_{\text{FPGA}}$ application executes within the secure world of TrustZone, while user applications ($AppX$) operate in the non-secure world. Inter-core communication was achieved using Xilinx Inter-Processor Interrupts (IPI) and shared memory. The secure reconfiguration workflow proceeds through the following coordinated steps among the system components:
\begin{enumerate}
     \item \textit{Request Initiation:} The $AppX$ application issues a request for FPGA acceleration to  $TA_{\text{FPGA}}$.
     \item \textit{Resource Allocation:} $TA_{\text{FPGA}}$ checks the availability of reconfigurable regions and confirms that the required vFPGA is available for allocation.
     \item \textit{Session Key Exchange:} To establish a secure communication channel, $TA_{\text{FPGA}}$ and $AppX$ exchange a session key. This key is used to protect the confidentiality and integrity of the partial bitstream. Here, AES is exchanged, and RSA is used for encapsulating the AES session keys. 
     \item \textit{Bitstream Preparation:} $AppX$ encrypts the partial bitstream through AES.%
     \item \textit{Bitstream Transfer:} $AppX$ sends the encrypted partial bitstream and notifies $TA_{\text{FPGA}}$ of the bitstream’s metadata, such as the intended location (vFPGA1 or vFPGA2) and its size over the secure channel.
     \item \textit{Decryption and Verification:} $TA_{\text{FPGA}}$ retrieves the encrypted partial bitstream from shared memory, decrypts, and verifies it.
    \item \textit{Configuration Instruction:} Once verified, $TA_{\text{FPGA}}$ provides the FPGA Shell with the decrypted partial bitstream, together with its intended location and size.
     \item \textit{Partial Reconfiguration:} The FPGA Shell configures the designated vFPGA region by streaming the partial bitstream through the internal configuration port (ICAP).
     \item \textit{Acknowledgment:} After successful configuration, the FPGA Shell acknowledges completion to $TA_{\text{FPGA}}$, which in turn issues a final acknowledgment to $AppX$.
\end{enumerate}
This workflow ensures end-to-end protection of partial bitstream confidentiality and integrity while maintaining isolation between the control logic and user applications.

To illustrate practical usage, we implemented vFPGAs on the same FPGA. The statistics for the allocated size of blocks provide an estimate of resource usage for the two implemented applications on the FPGA, based on Vivado's floorplanning estimator.
(1) \textit{vFPGA1:} A lightweight CNN accelerator for image classification, comprising a 3×3 convolution, quantization, ReLU activation, and pooling, processing a 6×6 input feature map. Resource usage includes 30 CLB LUTs, 30 LUTs as Logic, 32 CLB Registers, 32 Registers as flip-flops, and one F7 Muxes.
(2) \textit{vFPGA2:} A configurable shift circuit supporting left and right shift operations. Resource usage includes 2 CLB LUTs, 2 LUTs as Logic, 35 CLB Registers, 35 Registers as Flip Flops, 5 CARRY8s, one Block RAM Tile, and RAMB36/FIFO.

\begin{table}[h!]
\centering
\caption{Configuration times for different partial bitstreams.}
\label{tab:bitsize_perform}
\begin{tabular}{|c|>{\centering\arraybackslash}p{3cm}|>{\centering\arraybackslash}p{3cm}|}
\hline
\textbf{Module}    & \textbf{Mean Up. Time (ms)}& \textbf{Std. Dev. (ms)}  \\
\hline
vFPGA1          & 495.21                         & 8.64 \\
\hline
vFPGA2           & 528.21                         & 0.27 \\
\hline
\end{tabular}
\end{table}

The system was tested end-to-end by simulating a secure \textit{ground station to satellite} reconfiguration scenario. $AppX$ acted as the ground station, transmitting a partial bitstream and emulating a radio link. $TA_{\text{FPGA}}$ works as the receiver of the satellite. The update triggered secure boot procedures, cryptographic validation, and partial reconfiguration on the FPGA. Pre-emptive detection of hardware Trojans, especially those incorporating sensor or power-draining circuits, is essential prior to deploying configurations on the FPGA. 
Here we adopted the tool from ~\cite{la2020fpgadefender} to analyze and identify malicious circuits. Despite minimal cryptographic overhead and multi-core coordination, the reconfiguration latency remained acceptable, demonstrating the approach’s suitability for in-orbit reconfiguration scenarios. The reconfiguration time for each vFPGA is presented in Tables~\ref{tab:bitsize_perform}. The configuration times were measured by performing 25 trials for each partial bitstream. The sample standard deviation reflects the variability within these limited trials. While the example CNN is lightweight, the same methodology supports more complex accelerators.

\section{Research Outlook}
\label{sec:outlook}

As future missions demand greater adaptability, resilience, and assurance in adversarial or disconnected environments, several technical and systemic challenges persist. %

\subsection{Post-Quantum Cryptography}
Current secure boot and update mechanisms rely on classical public-key cryptography, such as RSA and ECDSA, which are increasingly vulnerable to quantum attacks~\cite{kumar2020post}. Given satellites’ long service lifetimes, integrating post-quantum cryptographic (PQC) schemes into SoC FPGA toolchains is essential. Algorithms like CRYSTALS-Dilithium and SPHINCS+ are advancing toward NIST standardization~\cite{NISTPQC}. However, practical implementation is still a major challenge. While proof-of-concept secure-boot demonstrations exist on terrestrial hardware, designing constant-time, radiation-tolerant PQC engines for space-grade, resource-constrained FPGAs remains largely unexplored. Optimizing PQC implementations for predictable timing and minimal memory use is also crucial to avoid side-channel vulnerabilities and ensure reliability in orbit.

\subsection{Autonomous Attestation in Disconnected Missions}
While secure boot establishes a root of trust at startup, long-duration deep-space missions require continuous trust evaluation without real-time ground oversight. Autonomous attestation mechanisms are needed to detect unexpected behavior, validate reconfiguration events, and respond to tampering evidence. Notably, the development of in-orbit autonomous attesters capable of continual self-measurement of the PL fabric under severe communication delays remains an open field. Future work should explore lightweight, self-verifying architectures to provide resilience in disconnected scenarios.

\subsection{Energy-Constrained Isolation Mechanisms}
Isolation mechanisms such as ARM TrustZone, AXI Firewalls, and SMMUs are critical for domain separation but may introduce latency and power overhead in energy-constrained platforms like CubeSats. Optimized isolation techniques at RTL and fabric level are needed, including logic-privilege tags, dynamic bus gating, and low-overhead privilege-checking circuits. Additionally, energy-adaptive security policies, wherein the isolation strength dynamically scales according to battery state and mission phase, remain largely unexplored. %

\subsection{Cross-Domain Hardware-Software Co-Design}
Securing AI-driven SoC FPGA satellites demands a holistic approach across the AI model lifecycle, from development to deployment and updates. Integrated toolchains must address FPGA synthesis, AI compiler output validation, secure key provisioning, and post-deployment telemetry, while accounting for aerospace constraints such as radiation, communication windows, and thermal budgets~\cite{popa2024architecture,cabrera2023errant}. Open-source toolchains could balance security and accessibility~\cite{ozdil2024model,pintos2024vitis}. Interdisciplinary collaboration is essential to develop standardized frameworks that meet mission requirements. A critical yet underexplored area is the integration of hardware-assisted side-channel monitors within the FPGA fabric itself, such as embedded glitch sensors and power anomaly detectors, to detect physical or electromagnetic probing attempts in orbit. %

\subsection{Federated Learning Security in Satellite Swarms}
Finally, emerging mission concepts envision large constellations of cooperating small satellites performing federated AI training and inference. Although secure aggregation protocols have been explored in terrestrial settings, federated learning security against poisoning and Byzantine faults across hundreds of Low Earth Orbit nodes remains unaddressed in hardware-rooted frameworks. Developing scalable, resilient architectures to enable trustworthy federated learning in space represents a promising frontier~\cite{singh2023ai, ModelPoisoning, li2023flairs}.

\section{Conclusion}
\label{sec:concl}
This paper addresses the pressing need for secure AI-enabled satellite platforms by presenting a comprehensive security framework, \ournamewspace tailored for SoC FPGAs. Recognizing the unique challenges posed by reconfigurable computing in space,\ournamewspace integrates multiple layers of protection. %
This holistic approach establishes a trusted chain from system initialization to ongoing operations, critical for maintaining mission integrity. By laying this security foundation, our work supports the safe and effective deployment of AI in space, enabling the next generation of intelligent satellite systems. As satellite technology evolves, future research must focus on adapting to new threats and constraints.

\section*{Acknowledgement}
Our research work was partially funded by Intel’s Scalable Assurance Program, Deutsche Forschungsgemeinschaft (DFG) – SFB 1119 – 236615297, the European Union under Horizon Europe Programme – Grant Agreement 101070537 – CrossCon, and the European Research Council under the ERC Programme - Grant 101055025 - HYDRANOS. This work does not in any way constitute an Intel endorsement of a product or supplier. Any opinions, findings, conclusions, or recommendations expressed herein are those of the authors and do not necessarily reflect those of Intel, the European Union, and the European Research Council.

\bibliographystyle{IEEEtran}
\bibliography{sample-base.bib}

\end{document}